\documentclass{iaus}
\usepackage{graphicx}
\usepackage{hyperref}

\title[From {\sc Hipparcos} to {\sc Gaia}] 
{From {\sc Hipparcos} to {\sc Gaia}}

\author[L.~Eyer et al.]   
{L.~Eyer$^{1}$,
P.~Dubath$^{1,2}$,
S.~Saesen$^{1}$,
D.W. Evans$^{3}$,
L. Wyrzykowski$^{3}$,
S. Hodgkin$^{3}$
and N.~Mowlavi$^{1,2}$}

\affiliation{$^1$Geneva Observatory, Department of Astronomy, University of Geneva, CH-1290 Sauverny, Switzerland \\  email: {\tt Laurent.Eyer@unige.ch} \\[\affilskip]
$^2$ISDC, Department of Astronomy, University of Geneva, CH-1290 Versoix, Switzerland\\[\affilskip]
$^3$Institute of Astronomy, Cambridge University, Cambridge, UK}

\pubyear{2012}
\volume{285}  
\pagerange{119--126}
\jname{New Horizons in Time Domain Astronomy}
\editors{E. Griffin, B. Hanisch \& R. Seaman, eds.}

\begin{document}

\maketitle

\begin{abstract}

The measurement of the positions, distances, motions and luminosities of stars represents the foundations of modern astronomical knowledge. Launched at the end of the eighties, the ESA {\sc Hipparcos} satellite was the first space mission dedicated to such measurements. Hipparcos improved position accuracies by a factor of 100 compared to typical ground-based results and provided astrometric and photometric multi-epoch observations of 118,000 stars over the entire sky. The impact of Hipparcos on astrophysics has been extremely valuable and diverse. Building on this important European success, the ESA {\sc Gaia} cornerstone mission promises an even more impressive advance. Compared to {\sc Hipparcos}, it will bring a gain of a factor 50 to 100 in position accuracy and of a factor of 10,000 in star number, collecting photometric, spectrophotometric and spectroscopic data for one billion celestial objects. During its 5-year flight, {\sc Gaia} will measure objects repeatedly, up to a few hundred times, providing an unprecedented database to study the variability of all types of celestial objects. {\sc Gaia} will bring outstanding contributions, directly or indirectly, to most fields of research in astrophysics, such as the study of our Galaxy and of its stellar constituents, the search for planets outside the solar system.

\end{abstract}

\firstsection 
\section{Introduction}

Time domain plays a central role in the {\sc Hipparcos} and {\sc Gaia} missions. Indeed, to determine parallaxes and proper motions of stars, repeated observations of positions are required.  The accuracy of ground-based astrometric measurements are limited because of the presence of atmospheric distortions, of telescope deformations and of difficulties related to the definition of the reference frame. Even if a set of distant stars are used, they may have streaming motions or residual parallaxes, leading to inaccuracies in parallax measurements.

Going into space, the deformations related to the presence of atmosphere and gravity are altogether avoided. The reference frame issue is solved in a similar manner for both {\sc Hipparcos} and {\sc Gaia}. The basic measurements are wide-angles between celestial targets. As the spacecraft is scanning the sky over an extended period of time, many wide-angles with many orientations are typically measured for a given target. The whole sphere is then reconstructed using some mathematical formulations, solving for the astrometric parameters of all targets globally. In this way, a full sky positional frame of unprecedented accuracy can be self calibrated (\cite[Lindegren \& Bastian~2011]{LindegrenBastian2011}).

Both of these missions are from the European Space Agency (ESA). Other astrometric missions were proposed, such as FAME and SIM from USA, but finally, they were not realized. There is also a Japanese astrometric project called JASMINE, the first step of which is a nano-satellite to be launched in 2013, cf. \url{http://www.jasmine-galaxy.org/index.html}. It should achieve {\sc Hipparcos} accuracies.

Although the astrometry measurement principles are similar for {\sc Hipparcos} and {\sc Gaia}, their scientific cases are different: the science case of {\sc Hipparcos} was mostly stellar, whereas the one of {\sc Gaia} is mostly oriented towards the Galaxy, aiming at studying its structure and formation history. However, {\sc Gaia} will also have a very significant impact on asteroid studies, stellar astronomy and fundamental physics.

Both missions also provide very valuable data sets for variability studies. The cadence of the observations is imposed by the satellite scanning laws optimized for best astrometric performances. It turns out however that the resulting time sampling is rather favorable for period determination  \cite[(Eyer \& Mignard 2005)]{EyerMignard2005}, as the one-day and one-year aliases (and the related ones) typically plaguing ground-based observations are absent and other aliases related to spacecraft characteristic periods are of rather low amplitude. In addition, both missions are performing a homogeneous whole sky survey with a single set of instruments and with fully standard data reduction procedures producing a very coherent data set.

Though the results of {\sc Hipparcos} were published in 1997, they are still unsurpassed in extent and quality. This is true for astrometry, but also for variability studies: {\sc Hipparcos} is still the only whole sky survey with a published systematic variability analysis.

\section{{\sc Hipparcos}}

The {\sc Hipparcos} mission collected observations for 118,000 stars up to magnitude Hp~=~12.4 grouping two main samples:

\begin{enumerate}

\item A whole sky survey sample of 52,000 stars brighter than 7.9 $V + 1.1 \sin(b)$ magnitude for blue stars with B-V $<$ 0.8 and brighter than 7.3 $V + 1.1 \sin(b)$ for redder stars with $B-V \geq 0.8$ (where $b$ is the galactic latitude).

\item A set of 66,000 objects selected on the basis on their scientific interest.

\end{enumerate}

The properties of these objects (position, proper motion, magnitude, color, spectral and variability types, radial velocities and multiplicity) were compiled and published in the {\sc Hipparcos} Input Catalogue (\cite[ESA 1992]{ESA1992}). There were also observational campaigns to collect complementary information during the mission. For example, the AAVSO (American Association of Variable Star Observers) measured the brightness of Long Period Variables in order to determine the correct exposure time for {\sc Hipparcos} for those brighter than Hp=12.4. Another campaign was done to recalibrate the Hp band by measuring R, N, C stars simultaneously from the ground and the space.

The satellite was launched from Kourou (French Guiana) on an Ariane-4 rocket in 1989.  The length of the mission was 3.3 years, during which a mean of 110 measurements per object were collected. The final results were published in a 17-volume catalogue in 1997 (\cite[ESA 1997]{ESA1997}). The mission reached its original goal to obtain astrometry at the milliarcsec level. In addition, the {\sc Hipparcos} Star Mapper provided measurements which lead to the Tycho and Tycho-2 catalogues (\cite[ESA 1997]{ESA1997}, \cite[Hoeg et al. 2000]{Hoegetal2000} resp.), containing astrometric and photometric ($B_T$ and $V_T$) measurements for 1 and 2.5 million stars, respectively. As expected, the Tycho epoch photometry in $B_T$ and $V_T$ has been more difficult to handle than the main mission photometry in the Hp band, and led to a relatively low number of variable stars, see \cite{Piquardetal2001}.
Ten years after the publication of the {\sc Hipparcos} catalogue, a new reduction of the astrometric data has been undertaken and published (\cite[van Leeuwen 2007]{vanLeeuwen2007}). The systematic effects are reduced for bright stars leading to errors consistent with the photon noise limits.

The project scientist of {\sc Hipparcos}, Michael Perryman, wrote two books (\cite[Perryman 2009]{Perryman2009}, \cite[Perryman 2010]{Perryman2010}). The first book summarizes and highlights the scientific harvest of {\sc Hipparcos} during the 10 years following the catalogue publication. The second is narrating the fantastic human, technological and scientific adventures of the {\sc Hipparcos} mission.

\subsection{A note on the Pleiades distance derived from {\sc Hipparcos}}
Distances of open clusters were published using {\sc Hipparcos} parallaxes. For the Pleiades, the distance obtained ($118.3 \pm 3.5$ pc, \cite[van Leeuwen 1999]{vanLeeuwen1999}) seemed in contradiction with the one obtained using the usual main-sequence fitting technique ($132 \pm 4$ pc, \cite[Meynet et  al. 1993]{Meynetetal1993}). Other independent determinations are in favor of the longer distance, e.g. \cite{Panetal2004}, \cite{Zwahlenetal2004}, \cite{Munarietal2004}, \cite{Soderblometal2005}. Although at the level of individual stars, the distances are compatible within the error estimations, the discrepancy becomes less acceptable when averaging.
The new {\sc Hipparcos} reduction by \cite{vanLeeuwen2007}  basically confirmed a small value of 120 $\pm$ 1.9 pc (\cite[van Leeuwen 2009]{vanLeeuwen2009}). {\sc Gaia} will close this debate on the distance of the Pleiades. Unfortunately for individual stars such as Atlas, {\sc Gaia} will not be able to provide a distance because of the brightness limit of its survey which is at V$\sim$6 mag.

\section{{\sc Gaia}}

The {\sc Gaia} program foresees to collect astrometric, photometric, spectrophotometric and spectroscopic measurements for one billion celestial objects with magnitudes between $V \sim 6$ and 20. The mission is designed to last 5 years, with a possible one-year extension, after a launch currently planned for 2013. The number of measurements expected after 5 years varies between 40 and 250 depending mostly on the target ecliptic latitude with 70 data points in average (the number is estimated taking into account dead-time estimation). Spectroscopic measurements with a resolution of 11,500 will be obtained for stars brighter than 16-17 mag.  The most recent performance estimates can be found on the following {\sc Gaia} webpage at ESA \url{http://www.rssd.esa.int/index.php?project=GAIA&page=Science_Performance}.  It is important to note that the astrometric accuracies are formulated as errors on the parallax at the end of the mission.

It has been realized  early in the preparation of the mission that the data processing is a challenge of the highest order. People interested in certain thematics formed working groups, which later evolved into Coordination Units (CU) (\cite[Mignard et al.~2008]{Mignardetal2008}). There are currently 8 Coordination Units and one of them, CU7, is responsible for all aspects related to the variability analysis of the data. The CU7 analyses start with calibrated data provided by other Coordination Units. Moreover variability and related time-series analyses have impacts on most Coordination Units. One additional Coordination Unit is currently being formed to develop the catalogue access interface for the scientific community. The {\sc Gaia} Data Processing and Analysis consortium (DPAC) now includes more than 500 scientists and software engineers.

Our estimations of the number of variable sources that {\sc Gaia} can detect range from 50 million to 150 million. This rather large range reveals limitations of our knowledge in the domain of variable phenomena, but it is also reflecting uncertainties on the ultimate precision reached once the satellite is in space. Some effort is currently being done to improve these estimations using results from other space missions, such as Kepler and CoRoT.

The diversity in variable phenomena is illustrated by the variability tree presented by \cite{EyerMowlavi2008}. {\sc Gaia} will detect most of the variability types from that tree.

\subsection{Variability Processing and analysis}

Within the {\sc Gaia} consortium, the CU7 group is dedicated to the variability analysis. It is in charge of analyzing all variable objects that appear above a certain variability threshold. Specific criteria are also defined do detect particular phenomena, such as short-period variables, solar-like variables, planetary transits and periodic small-amplitude variables. The specific behavior is taken into account to lower the variability thresholds set for general purpose. In other words, special variability detection can select variable stars that would be considered constant based on the general variability criteria.

Once the objects are considered as variable candidates, the variability is characterized by a certain number of parameters. Simple models such as a trend or periodic model are tested to see if those models can explain the variability. This characterization task produces a set of parameters, called attributes, that are used in the next step of the processing, the classification. The classification algorithms compute membership probabilities for each object on the basis of the specific values of the attributes. Supervised, unsupervised and semi-supervised methods are planned to be used. Different supervised methods have been evaluated using {\sc Hipparcos} data and random forest appears as one of the best algorithms in terms of overall performances (see \cite[Dubath et al. 2011]{Dubathetal2011}).

Once an object is classified and if it belongs to certain specific groups of variables, further specific analysis is carried out. Currently the different classes that will be treated into more detail are eclipsing binaries, solar-like variability (powered by magnetic fields), rotation-induced variable stars, RR Lyrae/Cepheid stars, long period variables, main-sequence pulsators, compact oscillators, pre-main sequence oscillators, microlensing, cataclysmic variables, rapid phases of stellar evolution and active galactic nuclei. In this process, some attributes are refined and additional parameters are computed, but another important activity is to evaluate the quality of the classification, where contamination and completeness are two fundamental criteria.

\subsection{Science alerts}

DPAC is preparing a science alerts system (see also \cite[Wyrzykowski \& Hodgkin~2011]{WyrzykowskiHodgkin2011}). The science alerts have been defined as events where `` the science data  would have little or no value without quick ground-based follow-up''. Several types of alerts are foreseen coming from the different instruments: astrometric alerts (include fast moving objects), photometric alerts and spectroscopic alerts.

Objects that could be detected with photometric alerts are for example supernovae (6,000 with 2,000 before peak); microlensing events  (1,000), with some having an astrometric signature; cataclysmic variables (novae, dwarf novae); eruptive stars (Be, RCrB stars, FU Ori); and possibly gamma ray bursts, including orphan afterglows. The detection of these photometric alerts will be performed on near-real-time {\sc Gaia} broad-filter photometry. The classification, however, will utilize all other data available immediately from {\sc Gaia}, including roughly calibrated low-resolution spectra. This will assure the best possible recognition of an alert. For supernovae, for example, it can provide a preliminary type and an estimate of the redshift.  {\sc Gaia} will also alert on Near Earth Objects and potential hazardous asteroids using its superb astrometry capability. One of the huge challenges for these science alerts is how to maximize completeness and avoid contamination from other uninteresting objects.

Since the data might be accumulated at the satellite location, there are time constraints which give a window within which alerts can be treated. This time window lies between a few hours and two days.
The alerts will not start from the first day of operation, but only after a 3-month verification phase. The goal is to verify the quality of the products of the alert pipeline and more precisely, the probabilities associated to the classification of these alerts. For the verification phase, a group of ground-based telescopes is being set-up. Furthermore, it is believed that scientific follow-up once the data is released publicly has to be organized in advance. Such structuring of the ground-based effort is being done now.

Also, with the development of large ground-based surveys, the specificities of the {\sc Gaia} alert system should be compared to other ones, such as those of PTF, CRTS, {\sc SkyMapper} and Pan-STARRS, and tuned to complement them as well as possible.

There has also been the idea to introduce a ``Watch List'', i.e., a list of targets suggested by interested scientists. Once these targets are behaving in an unusual way, an alert will be sent to the concerned scientist(s), so that a follow-up can be organized. This needs a good coordination among groups with different scientific interests, but is a good way to enhance the scientific return of {\sc Gaia} for selected objects.


\section{Conclusions}
The {\sc Hipparcos} mission has been a tremendous success. If the {\sc Gaia} mission results reach the current expectations, we can without any doubt say that its multi-epoch survey in astrometry, photometry, spectrophotometry and spectroscopy will have a considerable impact in astrophysics. We should also realize that {\sc Gaia} is not the only large-scale survey which is collecting data now, or which is planned. There are many other surveys like Pan-STARRS, LSST, OGLE, Sky-Mapper, PTF and CRTS to name a few. Also in combination with these other surveys, {\sc Gaia} will bring a significant contribution, because none of these surveys will reach the astrometric accuracy of {\sc Gaia}. In return, {\sc Gaia} will also benefit from these surveys, especially from photometric systems extending in the ultraviolet or infrared. In addition, a large public spectroscopic survey related to {\sc Gaia} on the VLT was selected by the European Southern Observatory. It starts in January 2012, and is expected to get 30 nights per semester for four years, with a fifth year subject to progress review.

\section*{Acknowledgement}
We would like to thank Prof. Michel Grenon, Prof. Joshua Bloom and Dr Pierre North for valuable discussions.
%

\end{document}